\documentclass[aps,prd,showkeys,showpacs,amssymb,cite,
amsfonts,epsf,preprintnumbers,nofootinbib,superscriptaddress]{revtex4}

\usepackage[dvips]{graphicx}
\usepackage{bm,latexsym,amsmath,amssymb,amsfonts,color}

\newcommand{\be}{\begin{equation}}
\newcommand{\ee}{\end{equation}}
\newcommand{\bear}{\begin{eqnarray}}
\newcommand{\eear}{\end{eqnarray}}
\newcommand{\ba}{\begin{array}}
\newcommand{\ea}{\end{array}}

\begin{document}

\title{Inflationary Tensor Perturbation in Eddington-inspired Born-Infeld gravity}

\author{Inyong Cho}
\email{iycho@seoultech.ac.kr}
\affiliation{Institute of Convergence Fundamental Studies \& School of Liberal Arts,
Seoul National University of Science and Technology, Seoul 139-743, Korea}
\author{Hyeong-Chan Kim}
\email{hckim@ut.ac.kr}
\affiliation{School of Liberal Arts and Sciences, Korea National University of Transportation, Chungju 380-702, Korea}

\begin{abstract}
We investigate the tensor perturbation in the 
inflation model
driven by a massive-scalar field in Eddington-inspired Born-Infeld gravity.
For short wave-length modes, the perturbation feature is very similar
to that of the usual chaotic inflation.
For long wave-length modes, the perturbation exhibits a peculiar rise
in the power spectrum which may leave a signature in the cosmic microwave background radiation.
\end{abstract}
\pacs{04.50.-h, 98.80.Cq, 98.80.-k}
\keywords{Inflation, Tensor Perturbation, Eddington-inspired Born-Infeld gravity}
\maketitle

\section{Introduction}
Recently, the evolution of the Universe driven by a massive scalar field was investigated in Ref.~\cite{Cho:2013pea}
in Eddington-inspired Born-Infeld (EiBI) gravity \cite{Banados:2010ix}.
The action for this model is given by
\begin{eqnarray}\label{action}
S_{{\rm EiBI}}=\frac{1}{\kappa}\int
d^4x\Big[~\sqrt{-|g_{\mu\nu}+\kappa
R_{\mu\nu}(\Gamma)|}-\lambda\sqrt{-|g_{\mu\nu}|}~\Big]+S_{\rm M}(g,\phi),
\end{eqnarray}
where $\lambda$  is a dimensionless parameter related with the
cosmological constant by $\Lambda = (\lambda -1)/\kappa$, and
$\kappa$ is the only additional parameter of the theory. In this
theory the metric $g_{\mu\nu}$ and the connection
$\Gamma_{\mu\nu}^{\rho}$ are treated as independent fields (Palatini
formalism). The Ricci tensor $R_{\mu\nu}(\Gamma)$ is evaluated
solely by the connection, and the matter field is coupled
only to the gravitational field $g_{\mu\nu}$.
The matter action is in the usual form used for the chaotic inflation model \cite{Linde:1983gd}
in general relativity (GR),
\be \label{S:chaotic}
S_{\rm M}(g,\phi) = \int d^4 x \sqrt{-|g_{\mu\nu}|}
\left[ -\frac12 g_{\mu\nu} \partial^\mu\phi \partial^\nu \phi -V(\phi) \right],
\qquad
V(\phi) = \frac{m^2}{2} \phi^2.
\ee

In EiBI gravity, there exists an upper bound in pressure
due to the square-root type of the action.
When the energy density is high,
the maximal pressure state (MPS) is achieved,
for which the Universe undergoes an exponential expansion
from a nonsingular initial state.
It was investigated that this MPS is the past attractor
from which all the classical evolution paths of the Universe originate.
Although the energy density is high in the MPS,
the curvature scale remains constant since the Hubble parameter becomes $H_{\rm MPS} \approx 2m/3$.
Therefore, quantum gravity is not necessary in describing the high-energy state
of the early universe.

The MPS is unstable under the global perturbation (zero-mode scalar perturbation)
and evolves to an inflationary attractor stage.
The succeeding inflation feature is the same with the ordinary chaotic inflation in GR,
but it is not chaotic at the high-energy state
because the pre-inflationary stage can have a finite low curvature.
Depending on the initial conditions, the evolution of the Universe
can acquire the 60 $e$-foldings in the late-time inflationary attractor period.
If the sufficient  $e$-foldings are not acquired in this period,
it must be complemented in the exponentially expanding period at the near-MPS
in order to solve the cosmological problems.

Once an inflation model has been introduced in EiBI gravity,
it is worthwhile to investigate the density perturbation. 
The density perturbation has been studied in the EiBI universe filled with perfect fluid
in Refs.~\cite{EscamillaRivera:2012vz,Avelino:2012ue,Lagos:2013aua,Yang:2013hsa}.
(Other work has been investigated in the cosmological and
astrophysical aspects in
Refs.~\cite{Cho:2012vg,Pani:2011mg,Pani:2012qb,DeFelice:2012hq,Avelino:2012ge,Avelino:2012qe,Casanellas:2011kf,
Liu:2012rc,Delsate:2012ky,Pani:2012qd,Cho:2013usa,Scargill:2012kg,Kim:2013nna,Kim:2013noa,Du:2014jka}.)
In particular, the very recent observational result of BICEP2
has put an increasing importance on the tensor-mode perturbation
in the inflationary scenario \cite{Ade:2014xna}.
In this paper, we investigate the tensor perturbation in the EiBI inflation model
introduced above.
We shall consider the perturbation analytically at the two stages:
``the near-MPS stage" which is described by the globally perturbed solution of the MPS,
and ``the attractor stage" which is similar to the ordinary chaotic inflation in GR.

\section{Formulation of Tensor Perturbation in EiBI gravity}
Equations of motion are derived by varying the action \eqref{action}
with respect to $g_{\mu\nu}$ and $\Gamma^\mu_{\alpha\beta}$,
and reduce to
\begin{align}
\frac{\sqrt{-|q|}}{\sqrt{-|g|}}~q^{\mu\nu}
& =\lambda g^{\mu\nu} -\kappa T^{\mu\nu},\label{eom1}\\
q_{\mu\nu} & = g_{\mu\nu}+\kappa R_{\mu\nu}, \label{eom2}
\end{align}
where $q_{\mu\nu}$ is newly defined auxiliary metric.
The EiBI theory is then described by the metric and the auxiliary metric
of which tensor perturbations are given by
\begin{align}
g_{\mu\nu}dx^\mu dx^\nu &=-a^2 d\eta^2 +a^2\left(\delta_{ij} +h_{ij}\right) dx^i dx^j ,\label{gmunu}\\
q_{\mu\nu}dx^\mu dx^\nu &=-X^2 d\eta^2 +Y^2 \left(\delta_{ij}+\gamma_{ij}\right)dx^i dx^j ,\label{eqmunu} \\
&=Y^2\left[ -d\tau^2 +\left(\delta_{ij}+\gamma_{ij}\right)dx^i dx^j \right],
\end{align}
where $\eta$ and $\tau$ are the conformal time for the metric
and for the auxiliary metric respectively.
We impose the transverse and traceless conditions
on both $h_{ij}$ and $\gamma_{ij}$, i.e.,
${\partial}_{i}h^{ij}={\partial}_{i}\gamma^{ij}=0$ and $h=\gamma=0$.
In what follows, we define $' \equiv d/d\eta$ and $\dot{} \equiv d/d\tau$.

From Eq.~\eqref{eom1}, one gets $\gamma_{ij} = h_{ij}$ \cite{EscamillaRivera:2012vz}.
The components of Eq.~\eqref{eom1} can be written as
\begin{eqnarray}\label{EOM11}
\frac{Y^3}{Xa^2} = \lambda +\kappa\rho, \qquad
\frac{XY}{a^2} = \lambda -\kappa p,
\end{eqnarray}
which give
\bear\label{XY}
X=(\lambda -\kappa p)^{3/4} (\lambda +\kappa\rho)^{-1/4} a,
\qquad
Y=(\lambda -\kappa p)^{1/4} (\lambda +\kappa\rho)^{1/4} a.
\eear

The tensor mode $h_{ij}$ possesses two degrees of freedom
corresponding to two polarizations of gravitational waves,
and can be expanded as
\be
h_{ij}(\eta,\vec{x}) =\sum_{\lambda = +,-}
\int \frac{d^3k}{(2\pi)^{3/2}} \;
h_{\lambda} (\eta,\vec{k}) \;
\epsilon^{\lambda}_{ij}(\vec{k}) \;
e^{i\vec k\cdot\vec x},
\ee
where $\epsilon^{\lambda}_{ij}$ represents the polarization tensor.
The transverse and traceless conditions reduce to
$k^i\epsilon^{\lambda}_{ij} =0$ and $\epsilon^{\lambda}=0$.
From Eq.~\eqref{eom2}, one gets the equation of motion for the perturbation,
\begin{eqnarray}\label{heq}
\frac{\kappa Y^2}{2X^2} h_{\lambda}''
+\frac{\kappa Y^2}{2X^2} \left(3\frac{Y'}{Y}-\frac{X'}{X}\right)h_{\lambda}'
+\frac{\kappa k^2}{2}h_{\lambda}=0.
\end{eqnarray}
Using $d\tau = (X/Y)d\eta$,
the above equation becomes
\be\label{heqintau}
\frac{\kappa}{2} \ddot h_\lambda + \kappa{\cal H}\dot h_\lambda
+\left[ \kappa (2{\cal H}^2 + \dot{\cal H} ) -Y^2 +a^2 +\frac{\kappa k^2}{2} \right] h_\lambda =0,
\ee
where ${\cal H} = \dot Y/Y$.

Now let us introduce the canonical field that describes the perturbation
by rescaling the mode function $h_\lambda$ by
\be\label{hmu}
h_\lambda(\tau,\vec{k}) \equiv f(\tau) \mu_\lambda(\tau,\vec{k}),
\ee
where $\mu_\lambda(\tau,\vec{k})$ is the canonical field and
$f(\tau)$ is the rescaling factor.
Plugging Eq.~\eqref{hmu} in Eq.~\eqref{heqintau},
one gets the equation of motion for $\mu_\lambda$,
\be\label{mueq}
\ddot\mu_\lambda + \Omega_k^2(\tau) \mu_\lambda =0,
\quad\mbox{where}\quad
\Omega_k^2(\tau) = k^2 + \frac{\ddot Y}{Y} +2\left( \frac{\dot Y}{Y}\right)^2
-\frac{2}{\kappa}(Y^2-a^2)
=k^2 -\frac{\ddot Y}{Y},
\ee
if $f$ satisfies
\be\label{hmu2}
\frac{\dot f}{f}+ \frac{\dot Y}{Y} = 0 \qquad\Rightarrow\qquad
f=\frac{f_0}{Y} \qquad\Rightarrow\qquad
h_\lambda = f_0 \frac{\mu_\lambda}{Y}.
\ee
Eq.~\eqref{mueq} is of the form which is derived from the action
for the canonical field,
\be\label{actionmu}
S_\mu = \int d\tau d^3x \left[ -\frac12 (\partial_\sigma \tilde\mu)^2 - U(\tilde\mu) \right],
\ee
where
\begin{align}\label{mutilde}
\tilde\mu_\lambda(\tau,\vec{x})= \int \frac{d^3k}{(2\pi)^{3/2}} \;\left[
\mu_\lambda(\tau,k)a_{\vec k} + \mu_\lambda^*(\tau,k)a_{-\vec k}^\dagger\right]
e^{i\vec k\cdot\vec x}.
\end{align}
From the action \eqref{actionmu} using Eq.~\eqref{mutilde},
the commutator relation provides the normalization condition for a given polarization mode,
\be\label{norm}
\left[ \tilde\mu_\lambda(\tau,\vec{x}) ,\Pi_{\tilde\mu_\lambda} (\tau,\vec{y}) \right]
= i \delta(\vec{x} - \vec{y})
\qquad\Rightarrow\qquad
\mu_\lambda\dot\mu_\lambda^{*} -\mu_\lambda^*\dot\mu_\lambda = i.
\ee
The derivation of $\mu_\lambda$ in Eq.~\eqref{hmu2} from Eq.~\eqref{heqintau}
is unique up to the constant $f_0$ which needs to be determined
from the reduction of the action.
Reducing the action \eqref{action} directly to Eq.~\eqref{actionmu},
we obtain $f_0=2$.

From the perturbation equation~\eqref{mueq},
there are two things to note.
First, the time is governed by $\tau$ which is not the conformal time of the metric
but of the auxiliary metric.
Second, the usual role of the scale factor $a$ in the perturbation theory in GR,
is now played by the scale factor $Y$ of the auxiliary metric in EiBI gravity.
As we shall see later,
at the attractor stage the differences disappear.
At the near-MPS stage, however, nontrivial differences appear
and the perturbation story is altered.

\section{Tensor perturbation in two stages}
The evolution of the Universe modeled by the action \eqref{action} was studied with the metric
\be
ds^2 = -dt^2 +a^2(t) d{\bf x}^2,
\ee
where $t$ is the cosmological time.
In the evolution of the Universe, it was found that
there exist two exponentially expanding stages.
The first one occurs near the maximal pressure state (we shall call this the {\it near-MPS stage}).
This stage is described by the globally perturbed background solutions from the maximal pressure state,
which was investigated in Ref.~\cite{Cho:2013pea}.
The second one occurs at the {\it attractor stage} in which the slow-roll conditions are satisfied.
This stage is very similar to the ordinary chaotic inflationary background.

In the cosmic microwave background radiation,
the scale-invariant region in the power spectrum corresponds to $\ell < 20$.
The perturbation for this region is produced during the inflating expansion
in which the $e$-folding value is $55 \sim 60$, measured back from the end of inflation.
When the attractor stage is long enough so that it can provide 60-$e$ foldings,
this spectrum fully comes from this stage.
If not, the scale-invariant perturbation should be produced during the near-MPS stage.

Although the scale factor $a$ exhibits the exponential expansion in both stages,
the evolutions of the scalar field $\phi$ in the two stages are quite different.
Therefore, the perturbations produced at two stages may differ.
Now, let us investigate the production of the perturbation at these two stages.

\subsection{Near-MPS stage}
The background fields $a$ and $\phi$ at the near-MPS stage
are described by the {\it maximal pressure solutions} plus
the globally (zero-mode scalar) perturbed solutions
discovered in Ref.~\cite{Cho:2013pea}.
The maximal pressure condition is given by
\be
\lambdabar -p = \lambdabar -\frac{\hat{\phi}^2}{2} +V(\phi) =0,
\ee
where $\lambdabar \equiv \lambda/\kappa$, $\hat{} \equiv d/dt$,
and in this paper we consider $\phi$
rolling on the left side of $V(\phi)$ initially.
Then the solution of $\phi$ for this condition is readily obtained,
and the corresponding solution for $a$ obtained by solving the Friedmann equation is
\begin{align}
\phi(t) &= \frac{\sqrt{2\lambdabar}}{m}\sinh(mt), \label{MPSphi}\\
a(t) &=a_0[U(\phi)]^{-2/3} = \frac{a_0}{(2\lambdabar)^{1/3}} \cosh^{-2/3}(mt)
\approx a_0 \left(\frac{2}{\lambdabar} \right)^{1/3} \; e^{2mt/3}, \label{MPSa}
\end{align}
where $U(\phi) \equiv \sqrt{2[V(\phi)+\lambdabar]}$.
The global perturbations were obtained in terms of the perturbation fields $h(t)$ and $\psi(t)$
which are defined by
\begin{align}
H &= \frac{\hat{a}}{a} = -\frac23 DU(\phi) \left[1+ h(t)\right], \label{GPH}\\
\hat\phi &=  U(\phi) \left[1+ \psi(t)\right], \label{GPphi}
\end{align}
where $D \equiv d/d\phi$ and $h(t), \psi(t) \ll 1$.
Note that when $h(t)=\psi(t) =0$ in these equations,
the relations for $H$ and $\hat{\phi}$ represent
simply the unperturbed background MPS.
The perturbed fields were obtained,
\begin{align}
h &= \psi_0 \left[-\frac{2}{3}+\sqrt{\frac{2}{3\kappa}}\frac{1}{DU(\phi)}\right]
\big[U(\phi)\big]^{-4/3}\; e^{t/t_c}, \label{GPh} \\
\psi &= \psi_0 \big[U(\phi)\big]^{-4/3} \; e^{t/t_c}, \label{GPpsi}
\end{align}
where $t_c =\sqrt{3\kappa/8}$.
As $t \to -\infty$, one gets
\begin{align}
DU(\phi) h &= \psi_0
 (2\lambdabar)^{-2/3} \left[ \sqrt{\frac{2}{3\kappa}} -\frac{2m}{3} \tanh(mt) \right]
 \cosh^{-4/3}(mt) \; e^{t/t_c}
 \propto
 e^{\left( 4m/3 + \sqrt{8/3\kappa} \right)t}, \label{Uh}\\
U(\phi) \psi &= \psi_0
 (2\lambdabar)^{-1/6} \cosh^{-1/3}(mt) \; e^{t/t_c}
 \propto
 e^{\left( m/3 + \sqrt{8/3\kappa} \right)t}.\label{Upsi}
\end{align}
Using the above results, we have following expansions,
\begin{align}
&\lambdabar-p = -\psi U^2 \left(1+\frac{1}{2}\psi\right), \label{lambda-p}\\
&\lambdabar +\rho = U^2 \left( 1+ \psi +\frac{1}{2}\psi^2\right). \label{lambda+rho}
\end{align}
Using Eqs.~\eqref{MPSa}, \eqref{GPpsi}, \eqref{lambda-p} and \eqref{lambda+rho},
one can get the following quantity in the lowest order,
\be
\frac{aY}{X} =a\sqrt{\frac{\lambdabar+\rho}{\lambdabar-p}}
\approx \frac{a}{\sqrt{-\psi}}
=\frac{a_0}{\sqrt{-\psi_0}}e^{-t/2t_c}
\equiv A_0e^{-t/2t_c}.
\ee
Then from the transformation between the cosmological time and the conformal time,
$d\tau = (X/aY)dt$, one can get the relation between two time coordinates,
\be\label{tauvst}
\tau = \int^t_{-\infty} \frac{X}{aY} dt'
\approx \frac{1}{A_0}  \int^t_{-\infty} e^{t'/2t_c} dt'
=\frac{2t_c}{A_0}e^{t/2t_c}.
\ee
The scale factor of the auxiliary metric, $Y(\tau)$,
is obtained in the lowest order of $\epsilon$,
\begin{align}
Y(\tau) &=\kappa^{1/2} (\lambdabar -p)^{1/4} (\lambdabar +\rho)^{1/4} a
\approx \kappa^{1/4} (-\psi)^{1/4} U a  \nonumber \\
&= \kappa^{1/4} (-\psi_0)^{1/4} U^{2/3}a e^{t/4t_c}
=  \kappa^{1/4} (-\psi_0)^{1/4} a_0 e^{t/4t_c} \nonumber \\
&= \frac{\kappa^{1/4}a_0^{3/2}}{\sqrt{2t_c}} \sqrt{\tau}
\equiv \tau_Y \sqrt{\tau}, \label{Ytau}
\end{align}
while the scale factor of the metric in Eq.~\eqref{MPSa} is given by
the transformation Eq.~\eqref{tauvst} as
\be\label{atauMPS}
a(\tau) \approx \frac{3\tau_m}{2m} \tau^{m\sqrt{2\kappa/3}},
\quad\mbox{where}\quad
\tau_m = \frac{2ma_0}{3} \left( \frac{2}{\lambdabar} \right)^{1/3}
\left( \frac{A_0}{2t_c} \right)^{m\sqrt{2\kappa/3}}.
\ee
Finally, the equation of motion \eqref{mueq} at the near-MPS stage becomes
\be\label{mueq2}
\ddot\mu_\lambda + \Omega_k^2(\tau) \mu_\lambda =
\ddot\mu_\lambda + \left( k^2 -\frac{\ddot Y}{Y} \right)  \mu_\lambda
\approx
\ddot\mu_\lambda + \left( k^2 +\frac{1}{4\tau^2} \right) \mu_\lambda  \approx 0,
\ee
of which solution is given by
\be\label{mutau}
\mu_\lambda(\tau) = \sqrt{\tau}\Big[ c_1J_0(k\tau) + c_2Y_0(k\tau) \Big].
\ee

Let us analyze the perturbation.
In GR, the perturbation equation is of the form,
\be\label{mueqGR}
\mu''_\lambda + \left( k^2 -\frac{a''}{a} \right)  \mu_\lambda =0,
\ee
and $a''/a \sim 2H^2a^2$ for de Sitter-like expansions.
Therefore, the relation $k^2 \sim a''/a$
is equivalent with the relation $\lambda_{\rm phys} = a/k \sim H^{-1}$
which stands for the horizon crossing of the mode.
When $k^2 \gg a''/a$, the mode function is a plane wave.
This corresponds to $\lambda_{\rm phys} \ll  H^{-1}$
which means the wave length is the subhorizon scale.
When $k^2 \ll a''/a$, the perturbation is superhorizon scale
and the mode function increases monotonically.

In EiBI gravity however, as it was mentioned earlier,
the role of $a$ and $\eta$ are replaced by $Y$ and $\tau$, respectively.
in the perturbation equation \eqref{mueq2}.
The relation $k^2 \sim \ddot{Y}/Y$ does not coincide with
the relation $\lambda_{\rm phys} \sim H^{-1}$;
$k^2 \sim \ddot{Y}/Y$ reduces to $k \sim 1/\tau$.
The behavior of the mode function is as following;

(i) When $k^2 \gg \ddot Y/Y$ (i.e., $k \gg 1/\tau$),
the mode function $\mu_\lambda$ becomes a plane wave,
as one can see from the asymptotic behavior of the Bessel functions in Eq.~\eqref{mutau},
or directly from the equation of motion.

(ii) When $k^2 \ll \ddot Y/Y$ (i.e., $k \ll 1/\tau$),
the mode function evolves monotonically.

\begin{figure}[btph]
\begin{center}
\includegraphics[width=.6\linewidth,origin=tl]{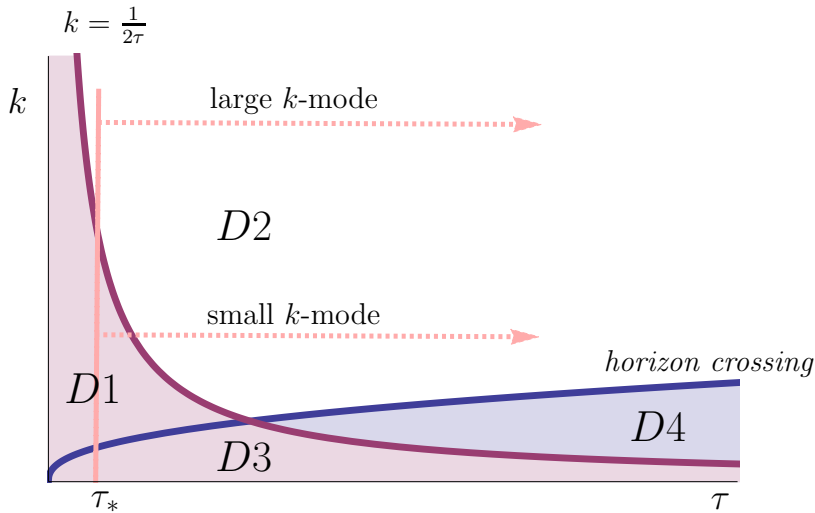}
\end{center}
\caption{Schematic plot of characteristic domains
at the near-MPS stage for the tensor perturbation.
Low $k$-modes can be initially formed in domain $D1$,
while high $k$-modes in $D2$.
In domains $D1$ and $D2$, the perturbation is in the subhorizon scale.
In domains $D2$ and $D4$, the mode function is a plane wave.
}
\label{fig:kvstau}
\end{figure}

\noindent At the near-MPS stage,
the background is almost de Sitter and
the horizon scale remains constant, $H^{-1} \approx 3/2m$.
The horizon crossing occurs when
\begin{align}
\lambda_{\rm phys} = \frac{a}{k} \sim H^{-1}
\qquad\Rightarrow\qquad
k \sim \frac{2}{3}ma
\approx \tau_m \tau^{m\sqrt{2\kappa/3}}. \label{k-hcrossing}
\end{align}
The mode is sub/superhorizon scale for $k \gtrless \tau_m \tau^{m\sqrt{2\kappa/3}}$.
The result \eqref{k-hcrossing}  does not coincide with $k \sim 1/\tau$,
so the behavior of the mode function is irrelevant to the horizon crossing.
(See Fig.~\ref{fig:kvstau}.)

Now let us discuss the production of the initial tensor perturbation at the near-MPS stage.
According to the pre-inflationary story investigated in Ref.~\cite{Cho:2013pea},
there is no obstacle to consider the Universe at $\tau =0$
for which the scalar field becomes $\phi \to -\infty$.
At this moment, the tensor perturbation $h_\lambda$ diverges logarithmically
because of $Y_0(k\tau)$.
(One cannot drop this term by setting $c_2=0$
because $\mu_\lambda$ is not normalizable in that case.)
However, it is not necessary to consider the production of the tensor perturbation at $\tau =0$.
Although the background curvature scale $H$ is finite,
the wave-length scale of the perturbation goes beyond the Planck scale,
for which the classical treatment of the perturbation is questioned.
In this sense, it is reasonable to consider the production of the perturbation
when the wave-length scale $\lambda_{\rm phys}$ is comparable to the Planck scale $l_{p}$,
\be\label{lplanck}
\lambda_{\rm phys} = \frac{a(\tau_*)}{k} \gtrsim l_p
\qquad\Rightarrow\qquad
\tau_* \gtrsim a^{-1} (kl_p) \approx \left( \frac{2kml_p}{3\tau_m}\right)^{\sqrt{3/2\kappa}/m} ,
\ee
where $\tau=\tau_*$ is the production moment of the perturbation.
When the initial perturbation is produced  at the near-MPS stage,
the perturbation solution is given by Eq.~\eqref{mutau}.
As usual, we impose the minimum-energy condition at $\tau_*$,
then the coefficients $c_1$ and $c_2$ in Eq.~\eqref{mutau} are fixed as following.
The energy for a given $k$ mode is given by
\be\label{EE}
E =\frac12 \left[|\dot \mu_\lambda|^2 + \left(k^2 +\frac1{4\tau^2} \right)|\mu_\lambda|^2\right].
\ee
From the mode solution~\eqref{mutau}, we get
\be\label{mu-smallk}
\dot \mu_\lambda(\tau) = \frac{1}{2\sqrt{\tau}} (c_1J_0+c_2Y_0)-k \sqrt{\tau} (c_1J_1+c_2Y_1).
\ee
The coefficients $c_1$ and $c_2$ are complex numbers,
\be
c_1 = c_1^{\rm Re} + ic_1^{\rm Im} \equiv c,
\qquad
c_2 = c_2^{\rm Re} + ic_2^{\rm Im} \equiv R-i \frac{\pi}{4c},
\ee
where $c$ and $R$ are real.
Here, we fixed one arbitrariness by imposing $c_1^{\rm Im}=0$,
and $c_2^{\rm Im}$ was determined from the normalization condition \eqref{norm}.
Then the energy \eqref{EE} becomes
\begin{align}\label{EE1}
8\tau E = \Big[cJ_0+RY_0 -2k\tau (cJ_1+RY_1)\Big]^2
+ (1+ 4k^2\tau^2)(cJ_0+RY_0)^2
+\left(\frac{\pi}{4c}\right)^2\Big[(Y_0-2k\tau Y_1)^2+ (1+4k^2\tau^2)Y_0^2 \Big] .
\end{align}

\subsubsection{High $k$-modes}
For high $k$-modes, the perturbation is produced in domain $D2$.
There, the mode function is oscillatory ($k\tau_* \gg 1$)
and the perturbation scale is subhorizon.
Then the energy becomes
\begin{align}
E & \approx 2k^2\tau_* \left[ c^2(J_0^2+J_1^2)  +
   R^2(Y_0^2+Y_1^2) + 2cR(J_0Y_0+J_1Y_1)  + \left(\frac{\pi}{4c}\right)^2 (Y_0^2+Y_1^2) \right]_{\tau=\tau_*}
   \nonumber \\
    & \approx \frac{k}{\pi} \left[c^2+R^2 + \left(\frac{\pi}{4c}\right)^2\right],
    \label{E:highk}
\end{align}
where, in the last step, we used the asymptotic formulae for $k\tau \gg 1$,
\be
J_0 = -Y_1 \approx \sqrt{\frac{2}{\pi k \tau} } \cos (k\tau -\frac{\pi}4), \qquad
Y_0 = J_1  \approx \sqrt{\frac{2}{\pi k \tau} } \sin (k\tau -\frac{\pi}4).
\ee
The energy~\eqref{E:highk} is minimized when $R=0$ and $c^2=\pi/4$,
which is exactly the same with the choice of the positive-energy mode
in the plane-wave type solution ($c_1=ic_2$).
Then the solution \eqref{mutau} for high $k$-modes becomes
\be\label{mu+MPS}
\mu_\lambda(\tau) = c\sqrt{\frac{2}{\pi k}} e^{i\pi/4} e^{-ik\tau}
= \pm \frac{1}{\sqrt{2k}}e^{i\pi/4}e^{-ik\tau}.
\ee

\subsubsection{Low $k$-modes}
\begin{figure}[btph]
\begin{center}
\includegraphics[width=.6\linewidth,origin=tl]{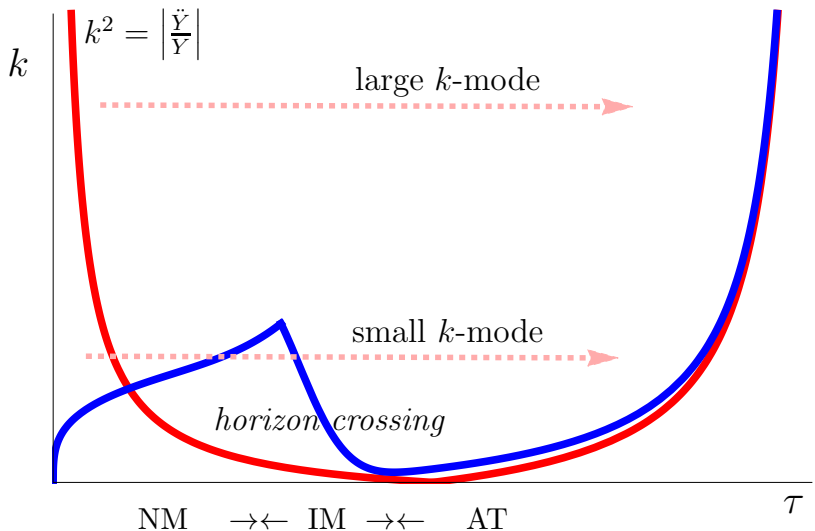}
\end{center}
\caption{Schematic plot of horizon crossing (blue) and $k^2$ vs. $|\ddot Y/Y|$ (red).
NM/IM/AT stand for near-MPS, intermediate, and attractor stages individually.
Above the blue line, the mode is subhorizon scale.
Above the red line, the mode is a plane wave.
The high $k$-mode remains inside the horizon as a plane wave till the attractor stage.
The low $k$-mode becomes a plane wave
after crossing the red line at early times
and maintains till the attractor stage.
It may cross the horizon twice in the mean time.
}
\label{fig:Yhorizon}
\end{figure}

For low $k$-modes, the initial perturbation is produced
at the nonoscillatory domain $D1$ ($k\tau_* < 1$).
Later, this perturbation evolves into the domain $D2$
and becomes oscillatory.
These modes can remain in $D2$ till the end of the near-MPS stage.
Lower $k$-modes can exit the horizon and enter into $D4$.
These modes re-enter the horizon at the following intermediate stage. (See. Fig.~\ref{fig:Yhorizon}.)
Even lower $k$-modes produced in $D3$ are not of cosmological interest.

The energy in Eq.~\eqref{EE1} can be rewritten as
\begin{eqnarray}
E &=&
\frac{\bar k^2}{8\tau}\left[ R^2(Y^2+Y_0^2) +
 2cR(JY +J_0 Y_0) +
 c^2(J^2+J_0^2) +
 \left(\frac{\pi}{4c}\right)^2(Y^2+Y_0^2) \right]_{\tau=\tau_*}, \label{E:gen}
\end{eqnarray}
where $J\equiv (J_0-2k\tau J_1)/\bar k$, $Y\equiv (Y_0-2k\tau Y_1)/\bar k$,
and $\bar k^2 \equiv 1+ 4k^2\tau^2$.
$E$ is minimized along the $R$-direction when
\be
R = - \frac{JY + J_0 Y_0}{Y^2+ Y_0^2} c.
\ee
Plugging this relation in Eq.~\eqref{E:gen},
we get the minimum energy when
\begin{eqnarray}\label{c-R}
c^2 &=&\frac{\pi}4 \frac{Y^2 +Y_0^2}{|JY_0-J_0Y|}
\qquad\Rightarrow\qquad
R=\mp \sqrt{\frac{\pi}{4}}\frac{JY + J_0 Y_0}{\sqrt{|JY_0-J_0Y|(Y^2 +Y_0^2)}} \;.
\end{eqnarray}
Note that $c$ and $R$ are determined at the production moment $\tau_*$
and thus they are functions of $k\tau_*$ only.
With these $c$ and $R$, the solution \eqref{mutau} for low $k$-modes becomes
\be\label{mu-MPS}
\mu_\lambda(\tau)  = \sqrt{\tau}\left[c J_0(k\tau)
+ \left( R-i\frac{\pi}{4c}\right)Y_0(k\tau) \right] .
\ee

\vspace{12pt}
In this section, we investigated the tensor perturbation at the near-MPS stage.
We obtained the general solution in Eq.~\eqref{mutau}.
In order to fix its coefficients,
we imposed the minimum-energy condition as an initial condition,
and then finally obtained the solutions \eqref{mu+MPS} and \eqref{mu-MPS}
for high and low $k$-modes.
Very low $k$-modes may exit the horizon at the near-MPS stage,
but these long wave-length modes are not of cosmological interests.
The others will stay inside the horizon till the end of the near-MPS stage,
and will evolve into the later stages.
These scales exit the horizon at the later stage,
and may leave a cosmological signature.

\subsection{Attractor stage}
After the near-MPS stage, the background universe evolves
into the intermediate stage at which only the first slow-roll condition is satisfied.
Afterwards, the Universe enters the attractor stage at which
both of the first and the second slow-roll conditions are satisfied.
The background evolution was found \cite{Cho:2013pea} to be very similar to that
in the usual chaotic inflation in GR.
In this subsection, we investigate the tensor perturbation at the attractor stage.

At the attractor stage,
the scalar field and scale factor are given by
\be\label{slowsol}
\phi(t) \approx \phi_i +\sqrt{\frac{2}{3}}m t,
\qquad
a(t) \approx a_i\; e^{[\phi_i^2-\phi^2(t)]/4},
\ee
where $\phi_i <0$ is the value of the scalar field in the beginning of the attractor.
For $N\sim 60$ $e$-foldings, $|\phi_i| \gtrsim 15$ is required.
From observational data, $m \sim 10^{-5}$ for the standard inflationary model.

At the early stage of the attractor, $m^2t^2 \ll mt$.
Then the scale factor is further approximated as
\be\label{aATT}
a(t) \approx a_i e^{-\phi_i mt/\sqrt{6} - m^2t^2/6}
\approx a_i e^{-\phi_i mt/\sqrt{6}}.
\ee
If the first slow-roll condition is applied, we have
\begin{align}
\rho &=\frac{\hat\phi^2}{2} +\frac{m^2}{2}\phi^2 \approx \frac{m^2}{2}\phi^2
= \frac{m^2}{2} \left( \phi_i +\sqrt{\frac{2}{3}} mt \right)^2
\approx  \frac{m^2}{2} \left( \phi_i^2 +\sqrt{\frac{8}{3}} \phi_i mt \right), \label{rhoATT}\\
p &= \frac{\hat\phi^2}{2} -\frac{m^2}{2}\phi^2 \approx -\frac{m^2}{2}\phi^2
\approx  -\frac{m^2}{2} \left( \phi_i^2 +\sqrt{\frac{8}{3}} \phi_i mt \right) \approx -\rho, \label{pATT}
\end{align}
which gives $X/Y = \sqrt{(\lambdabar-p)/(\lambdabar+\rho)} \approx 1$.
The time coordinates are transformed by
\be\label{tau-t}
d\tau = \frac{X}{Y}d\eta = \frac{X}{aY}dt \approx \frac{dt}{a}
\quad\Rightarrow\quad
\int_{\tau_i}^\tau d\tau' = \int_0^t \frac{dt'}{a(t')}
\quad\Rightarrow\quad
\tau-\tau_i = \frac{\sqrt{6}}{\phi_im} \left( \frac{1}{a} -\frac{1}{a_i} \right),
\ee
where we assumed that the attractor stage begins at $t=0$ and $\tau=\tau_i$.
One needs to be a bit cautious in setting the range of the time coordinates
because both $t$ and $\tau$ were used in the precedent near-MPS stage
in which the origin corresponded to $t\to -\infty$ and $\tau =0$.
Therefore, one needs $\tau>0$ (so $\tau_i >0$) for the attractor stage.
Setting $t=0$ for the beginning of the attractor stage fixes
the arbitrariness of the scale factor, $a(t=0)=a_i$.
From Eq.~\eqref{tau-t}, the scale factor can be obtained as
\be\label{aATT}
a(\tau) = \frac{a_i(\tau_i-\tau_0)}{\tau-\tau_0},
\qquad
\tau_0 \equiv \tau(t\to\infty) = \tau_i - \frac{\sqrt{6}}{\phi_im a_i}.
\ee
From Eqs.~\eqref{rhoATT} and \eqref{pATT}, we get
\be\label{YATT}
Y =\kappa^{1/2} (\lambdabar -p)^{1/4} (\lambdabar +\rho)^{1/4} a
\approx \kappa^{1/2}\sqrt{\lambdabar +\frac{1}{2}m^2\phi_i^2
+ 2m^2\log \left( \frac{\tau-\tau_i}{\tau_i -\tau_0} +1 \right)  } \;a
\equiv Y_0 a.
\ee
In the most of the attractor stage, the time dependence of $Y_0$
is subdominant to that of $a$.
Therefore, it can be regarded almost as constant,
\be
\frac{\ddot Y}{Y} = \frac{\ddot Y_0}{Y_0} +2 \frac{\dot Y_0\dot a}{Y_0a}
+\frac{\ddot a}{a} \approx \frac{\ddot a}{a}.
\ee
With this result, comparing  Eqs.~\eqref{mueq} and \eqref{mueqGR},
the perturbation behavior is the same with that of the ordinary chaotic inflation.
[Note from Eq.~\eqref{tau-t} that $d\tau =d\eta$ at the attractor stage.]
The perturbation equation \eqref{mueq} becomes 
\be\label{mueqATT}
\ddot\mu_\lambda + \left[ k^2 -\frac{2}{(\tau-\tau_0)^2} \right] \mu_\lambda  \approx 0,
\ee
of which solution is given by
\be\label{muATT}
\mu_\lambda(\tau)  = A_1 \left[ \cos k(\tau-\tau_0) - \frac{\sin k(\tau-\tau_0)}{k(\tau-\tau_0)} \right]
+A_2 \left[ \sin k(\tau-\tau_0) + \frac{\cos k(\tau-\tau_0)}{k(\tau-\tau_0)} \right].
\ee
The normalization condition~\eqref{norm} gives
\be\label{norm3}
A_1A_2^* - A_1^*A_2 = \frac{i}{k}.
\ee

\subsubsection{Subhorizon modes}
At the attractor stage, the horizon crossing occurs at
\be
\lambda_{\rm phys} =\frac{a}{k} \sim H^{-1}
\qquad\Rightarrow\qquad
|k(\tau-\tau_0)| \sim 1.
\ee
When the perturbation modes are subhorizon scale, $|k(\tau-\tau_0)| \gg 1$,
the solution \eqref{muATT} becomes
\be\label{musubATT}
\mu_\lambda(\tau)  \approx A_1 \cos k(\tau-\tau_0)  +A_2 \sin k(\tau-\tau_0),
\ee
which indicates a plane wave.
This type of subhorizon solution represents either

(i) the perturbation produced at the attractor stage
when the wave-length scale is comparable to the Planck scale, $\lambda_{\rm phys}\sim l_p$,
or

(ii) the continuation of the high $k$-mode perturbation \eqref{mu+MPS}
produced at the near-MPS stage.

\noindent In both cases, the minimum-energy condition
selects the positive-energy mode, $A_1=iA_2$,
\be\label{mu+ATT}
\mu_\lambda(\tau)
= \frac{1}{\sqrt{2k}} e^{i\theta} e^{-ik(\tau-\tau_0)},
\ee
where we used the normalization condition \eqref{norm3}
for the positive-energy mode, $(A_1^{\rm Re})^2+(A_2^{\rm Re})^2 =1/2k$.
We believe that this solution represents the high $k$-mode solution \eqref{mu+MPS}
which was produced at the near-MPS stage
and was continued through the intermediate stage remaining inside the horizon.
[This solution is identified with the one in Eq.~\eqref{mu+MPS}
when $\theta +k\tau_0 = \tan^{-1}(-A_2^{\rm Re}/A_1^{\rm Re}) =\pi/4$.]

\subsubsection{Superhorizon modes}
As time elapses, the physical wave-length scale grows,
exits the horizon, and becomes superhorizon scale in the end.
For the superhorizon and the horizon-crossing scales, $|k(\tau-\tau_0)| \lesssim 1$,
the second and fourth terms in Eq.~\eqref{muATT} are comparable, or dominant to
the other terms. Let us keep all the terms.
The relation between the coefficients, $A_1=iA_2$, and the normalization condition
maintain from the subhorizon scales.
Then the solution \eqref{muATT} becomes
\be\label{musuper+ATT}
\mu_\lambda(\tau)  = A_1 \left\{ \left[ \cos k(\tau-\tau_0)
- \frac{\sin k(\tau-\tau_0)}{k(\tau-\tau_0)} \right]
+i \left[ \sin k(\tau-\tau_0)
+ \frac{\cos k(\tau-\tau_0)}{k(\tau-\tau_0)} \right]\right\}
\quad\Rightarrow\quad
\mu_\lambda\mu_\lambda^*
=\frac{1}{2k} \left[ 1 + \frac{1}{k^2(\tau-\tau_0)^2} \right].
\ee

Now we can evaluate the tensor power spectrum for the perturbation
produced at the attractor stage,
or for the high $k$-mode perturbation
produced at the near-MPS stage,
\begin{align}
{\cal P}_{\rm T}(k)
&= \frac{k^3}{2\pi^2} P_h(k) = \frac{k^3}{2\pi^2} h_\lambda^2
=  \frac{k^3}{2\pi^2} \left( f_0 \frac{\mu_\lambda}{Y} \right)^2  \nonumber \\
&= \frac{ 1 + k^2(\tau-\tau_0)^2 }{\pi^2a_i^2(\tau_i-\tau_0)^2(\lambda+\kappa m^2\phi_i^2/2)}
\quad\to\quad
\frac{1}{\pi^2a_i^2(\tau_i-\tau_0)^2(\lambda+\kappa m^2\phi_i^2/2)}
\quad\mbox{(as $\tau\to\tau_0$)}.\label{PSATT}
\end{align}
The result is the same with that of the usual chaotic inflation
except a small EiBI correction, $\kappa m^2\phi_i^2/2$,
when there is no cosmological constant ($\lambda=1$).
The power spectrum is scale invariant.

\vspace{12pt}
As a whole, the perturbation story produced
and exited at the attractor stage
(or high $k$-modes continued from the near MPS stage)
is very similar to that in GR.
This is because $Y = Y_0a \propto a$.
The difference appears through the coefficient $Y_0$
which includes the model parameters $\lambda$ and $\kappa$.

\section{Solution Matching and Power Spectrum for low $k$-modes}\label{sec:SolMat}
In the previous section, we obtain the tensor power spectrum
for the perturbation of high $k$-modes produced at the near-MPS stage,
or for that produced at the attractor stage.
Those perturbations were the positive-energy mode of the plane wave
from the initial (positive-energy) condition imposed at the production.

However, for the low $k$-modes produced at the near-MPS stage,
the mode function \eqref{mu-MPS} was not a plane wave
at the initial production moment
although the same initial condition was imposed.
For these modes, we need to approach in a different manner
in order to evaluate the power spectrum.

The power spectrum is evaluated by the mode solution \eqref{muATT}
of the attractor stage.
What we do not know is the coefficients $A_1$ and $A_2$
which are determined from the initial condition.
For low $k$-modes, the initial condition is imposed at the near-MPS stage
and fixes the coefficients.
The mode solution is given by Eq.~\eqref{mu-MPS}
which is different from the attractor solution \eqref{muATT}.
Therefore, we need the matching between two solutions
in order to determine $A_1$ and $A_2$.
(For high $k$-modes, the near-MPS solution \eqref{mu+MPS}
was identical with the attractor solution \eqref{mu+ATT}.
Therefore, there was not need for matching.)
Between the near-MPS and the attractor stages, however,
there is the intermediate stage
for which we do not know the background analytically.
In order to overcome this difficulty,
we assume that the perturbation evolves adiabatically,
and that there exists an adiabatic period which covers
the late part of the near-MPS stage, the intermediate stage,
and the early part of the attractor stage.
(See Fig.~2.)
In this period, for the perturbation equation
\be
\ddot\mu_\lambda + \Omega_k^2(\tau) \mu_\lambda = 0,
\ee
the solution is given by the WKB approximation in general,
\be\label{sol:WKB}
\mu_{\rm WKB}(\tau) = \frac{b_1}{\sqrt{2\Omega_k(\tau)}}
    \exp\left[ i \int^\tau \Omega_k(\tau') d\tau'\right]
+ \frac{b_2}{\sqrt{2\Omega_k(\tau)}}
    \exp\left[ -i \int^\tau \Omega_k(\tau') d\tau'\right].
\ee
For this WKB solution to be valid,
the adiabatic condition needs to be satisfied,
\be\label{ADepsilon}
\epsilon \equiv \Omega_k^{-3}\left| \frac{d\Omega_k^2}{d\tau}\right| \ll 1.
\ee
The solution $\mu_{\rm WKB}$ is supposed to match with the near-MPS solution
at $\tau=\tau_1$,
\be\label{muMPSmt}
\mu_{\rm MPS}(\tau) = \sqrt{\tau}\Big[ c_1J_0(k\tau) + c_2Y_0(k\tau) \Big],
\ee
and with the attractor solution at $\tau=\tau_2$,
\begin{align}
\mu_{\rm ATT}(\tau)  &= A_1 \left[ \cos k(\tau-\tau_0) - \frac{\sin k(\tau-\tau_0)}{k(\tau-\tau_0)} \right]
+A_2 \left[ \sin k(\tau-\tau_0) + \frac{\cos k(\tau-\tau_0)}{k(\tau-\tau_0)} \right] \nonumber \\
&= A_1' \left[1+ \frac{i}{k (\tau-\tau_0)}\right] e^{ik(\tau-\tau_0)}
+A_2'  \left[1- \frac{i}{k (\tau-\tau_0)}\right] e^{-ik(\tau-\tau_0)},\label{muATTmt2}
\end{align}
where $A_1'= (A_1-i A_2)/2$ and $A_2' =(A_1+i A_2)/2$.

\begin{figure}[btph]
\begin{center}
\includegraphics[width=.6\linewidth,origin=tl]{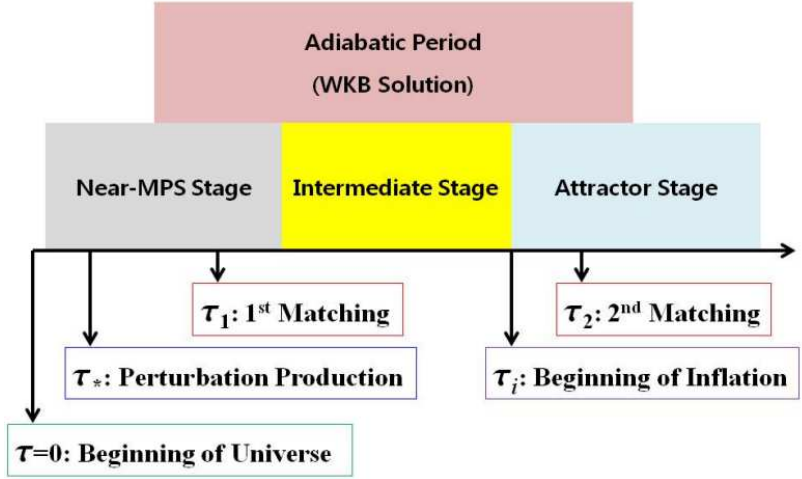}
\end{center}
\caption{Schematic plot of three stages and the adiabatic period.
Several important moments are marked.
}
\label{fig:diag}
\end{figure}

\subsection{Matching of $\mu_{\rm MPS}$ and $\mu_{\rm WKB}$}
At the near-MPS stage, the low $k$-mode solution is produced in $D1$ in Fig.~1,
where $k\tau_* <1$.
The coefficients $c_1$ and $c_2$ in Eq.~\eqref{muMPSmt} are determined as in Eq.~\eqref{mu-MPS}.
Later on, the perturbation evolves into $D2$.
At the near-MPS stage, the adiabatic condition becomes,
$\epsilon \ll  1 \to k\tau \gg 1$.
At $\tau =\tau_1 \gg 1/k$, the mode function becomes a plane wave
and can be approximated as
\begin{align}\label{muMPSmt2}
\mu_{\rm MPS}(\tau_1) &= \sqrt{\frac{2k}{\pi}}
\Big[ c_1 \cos\left(k\tau_1 -\frac{\pi}{4} \right)
+c_2 \sin\left(k\tau_1 -\frac{\pi}{4} \right) \Big].
\end{align}
At $\tau=\tau_1$, the WKB solution and its derivative become
\begin{align}
\mu_{\rm WKB}(\tau_1) &= \frac{b_1}{\sqrt{2\Omega_k(\tau_1)}}
    \exp\left[ i \int^{\tau_1} \Omega_k(\tau') d\tau'\right]
+ \frac{b_2}{\sqrt{2\Omega_k(\tau_1)}}
    \exp\left[ -i \int^{\tau_1} \Omega_k(\tau') d\tau'\right]
    \equiv \frac{b_1+b_2}{\sqrt{2\Omega_1}},\\
\dot\mu_{\rm WKB}(\tau_1) &= \sqrt{\frac{\Omega_1}{2}}
\left[ \left(i-\frac{\dot\Omega_1}{2\Omega_1^2} \right)b_1
- \left(i+\frac{\dot\Omega_1}{2\Omega_1^2} \right)b_2 \right]
\approx i \sqrt{\frac{\Omega_1}{2}}  (b_1-b_2),
\end{align}
where the integration factor was absorbed to the coefficients,
$\Omega_1 \equiv\Omega_k(\tau_1)$,
and we used $\dot\Omega_1/2\Omega_1^2 = \epsilon/4 \ll 1$.
Matching the solutions,
$\mu_{\rm MPS}(\tau_1)=\mu_{\rm WKB}(\tau_1)$ and
$\dot\mu_{\rm MPS}(\tau_1)=\dot\mu_{\rm WKB}(\tau_1)$,
gives
\be
b_{1,2} \approx \frac{c_1 \mp i c_2}{\sqrt{\pi}} \,e^{ \pm i \Phi_k},
\quad\mbox{where}\quad
\Phi_k = k \tau_1 - \frac{\pi}{4}.
\ee
Here, we used $\Omega_1^2 = k^2 +1/4\tau_1^2 \approx k^2$ since $k\tau_1 \gg 1$.

\subsection{Matching of $\mu_{\rm WKB}$ and $\mu_{\rm ATT}$}
At $\tau=\tau_2$, the WKB solution and its derivative become
\begin{align}
\mu_{\rm WKB}(\tau_2) &= \frac{b_1 e^{i \Psi} + b_2 e^{-i \Psi}}{\sqrt{2\Omega_2}}
	\approx \sqrt{\frac{2}{\pi\Omega_2}} \Big[ c_1 \cos (\Phi_k+\Psi) +c_2 \sin (\Phi_k+\Psi) \Big],\label{muWKBmt2}\\
\dot \mu_{\rm WKB}(\tau_2) &= i \sqrt{\frac{\Omega_2}{2}} (b_1 e^{i \Psi} - b_2 e^{-i \Psi})
\approx \sqrt{\frac{2\Omega_2}{\pi}} \Big[ -c_1 \sin (\Phi_k+\Psi)+ c_2 \cos (\Phi_k+\Psi) \Big],
\end{align}
where $\Psi \equiv \int^{\tau_2}\Omega_k(\tau')d\tau'$.
Using the solutions \eqref{muATTmt2} and \eqref{muWKBmt2},
the matching
$\mu_{\rm ATT}(\tau_2)=\mu_{\rm WKB}(\tau_2)$ and
$\dot\mu_{\rm ATT}(\tau_2)=\dot\mu_{\rm WKB}(\tau_2)$,
gives
\begin{align}
A_1' &= \frac{e^{-ik \delta \tau_2}}{2}\left[
	\left(1-\frac{i}{k \delta \tau_2} -\frac{1}{k^2 \delta \tau_2^2}\right)
	\mu_{\rm WKB}(\tau_2) -\frac{i}{k} \left(1-\frac{i}{k \delta \tau_2} \right)
	\dot\mu_{\rm WKB}(\tau_2) \right]
\approx \frac{e^{-ik \delta \tau_2}}{2}
\left[ \mu_{\rm WKB}(\tau_2) -\frac{i}{k} \dot\mu_{\rm WKB}(\tau_2) \right]   , \label{A1'}\\
A_2' &= \frac{e^{ik \delta \tau_2}}{2}\left[
	\left(1+\frac{i}{k \delta \tau_2} -\frac{1}{k^2 \delta \tau_2^2}\right)
	\mu_{\rm WKB}(\tau_2)
	+\frac{i}{k} \left(1+\frac{i}{k \delta \tau_2} \right)
	\dot\mu_{\rm WKB}(\tau_2) \right]
\approx \frac{e^{ik \delta \tau_2}}{2}
\left[ \mu_{\rm WKB}(\tau_2) +\frac{i}{k} \dot\mu_{\rm WKB}(\tau_2) \right] , \label{A2'}
\end{align}
where $\delta \tau_2 = \tau_2-\tau_0$
and the adiabatic condition at the attractor stage,
$\epsilon \ll 1 \to k\delta\tau_2 \gg 1$, was used.

\subsection{Power Spectrum}
The power spectrum is evaluated at the end of the attractor stage
($|k\delta\tau| \equiv |k(\tau-\tau_0)| \ll 1$).
The mode function \eqref{muATTmt2} in this limit becomes
\be
\mu_{\rm ATT}(\tau) \approx
i\frac{A_1'-A_2'}{k(\tau-\tau_0)}.
\ee
From Eqs.~\eqref{muWKBmt2}-\eqref{A2'}, we get
\begin{align}
A_1'-A_2' &\approx -i \sqrt{\frac{2}{\pi k}}
\left\{ \sin(k\delta\tau_2)\Big[c_1\cos(\Phi+\Psi) +c_2\sin(\Phi+\Psi) \Big]
+ \cos(k\delta\tau_2)\Big[-c_1\sin(\Phi+\Psi) +c_2\cos(\Phi+\Psi)\Big] \right\} \nonumber \\
&= -i \sqrt{\frac{2(c_1^2+c_2^2)}{\pi k}} \; \cos(\Upsilon +\Phi +\Psi -k\delta\tau_2),
\end{align}
where we used $\Omega_2^2 = k^2 -2/(\tau_2-\tau_0)^2 \approx k^2$
in the approximation since $|k(\tau_2-\tau_0)| \gg 1$,
and $\Upsilon\equiv \tan^{-1}(c_1/c_2)$.
At the end of the attractor stage ($|k(\tau-\tau_0)| \ll 1$),
we then get
\be
|\mu_{\rm ATT}|^2 \approx
\frac{|A_1'-A_2'|^2}{k^2(\tau-\tau_0)^2}
\approx \frac{2(c_1^2+c_2^2)}{\pi k^3 (\tau-\tau_0)^2}
\cos^2(\Upsilon +\Phi +\Psi -k\delta\tau_2)
=  \frac{c^2+R^2 + \pi^2/16c^2}{\pi k^3 (\tau-\tau_0)^2},
\ee
where the last result comes after averaging over several wave lengths.
The power spectrum evaluated at the end of the attractor stage becomes
\begin{align}
{\cal P}_{\rm T}(k)
&= \frac{k^3}{2\pi^2} \left( f_0 \frac{\mu_{\rm ATT}}{Y} \right)^2
\approx \frac{k^3f_0^2(\tau-\tau_0)^2 |\mu_{\rm ATT}|^2}{2\pi^2a_i^2(\tau_i-\tau_0)^2(\lambda+\kappa m^2\phi_i^2/2)}
\approx \frac{2(c^2+R^2 + \pi^2/16c^2)}{\pi^3 a_i^2(\tau_i-\tau_0)^2(\lambda+\kappa m^2\phi_i^2/2)}.\label{PS}
\end{align}
This power spectrum is $k$-dependent via $c(k\tau_*)$ and $R(k\tau_*)$
obtained in Eq.~\eqref{c-R}.
This results reduces to the power spectrum in Eq.~\eqref{PSATT}
in the high $k$-limit.
The power spectrum is insensitive to the matching time $\tau_1$ and $\tau_2$
(obtained in Appendix).
In Fig.~\ref{fig:PS}, the power spectrum ${\cal P}(T)$ is plotted.

From Fig.~\ref{fig:PS}, we can observe that the power spectrum is scale invariant
for $k \gtrsim 1/\tau_*$: this result is the same with that obtained in Eq.~\eqref{PSATT}
for the perturbation at the attractor stage.
However, the scale invariance is broken for the very low values of $k$.
This perturbation pattern produced in $D1$ is a distinct feature
from that of the ordinary inflationary model, and may leave a signature
in the cosmic microwave background.
[For the scale-invariant modes ($k \gtrsim 1/\tau_*$) in this figure,
the initial perturbation is formed in $D2$ in which the mode function is
the asymptotic plane-wave type of the Bessel function.
For the broken scale-invariant modes ($k \lesssim 1/\tau_*$),
the initial perturbation is formed in $D1$
in which the mode function has not been relaxed to
the asymptotic plane-wave type of the Bessel function.]

\begin{figure}[btph]
\begin{center}
\includegraphics[width=.5\linewidth,origin=tl]{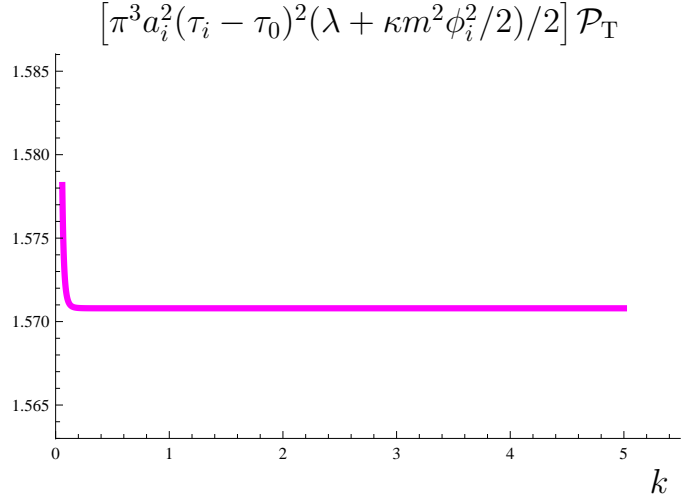}
\end{center}
\caption{Plot of $\left[ \pi^3 a_i^2(\tau_i-\tau_0)^2(\lambda+\kappa m^2\phi_i^2/2)/2 \right]{\cal P}_{\rm T}(k)$
for $a_0=-\psi_0=1$, $\lambda =1$, $m=10^{-4}$, $\kappa=10^{-4}$, and $\tau_*=13.68$.
}
\label{fig:PS}
\end{figure}

\vspace{12pt}
In summary, the perturbation considered in this section is for low $k$-modes.

(i) The initial perturbation is produced at $\tau_*$ in the domain $D1$ in Fig.~1.
Since it is in the domain of $k < 1/2\tau$,
the mode has not been fully relaxed to the asymptotic
plane-wave behavior of the Bessel function.
The mode solution is given by Eq.~\eqref{mu-smallk}
with the coefficients $(c_1,c_2)$ given by Eq.~\eqref{c-R}.

(ii) The perturbation produced as (i) evolves into the adiabatic period later on,
and the solution matches with the WKB solution \eqref{sol:WKB} at $\tau = \tau_1 > \tau_*$.
The coefficients $(b_1,b_2)$ of the WKB solution are then determined
from matching with the near-MPS solution at $\tau_1$.
At the near-MPS stage, the adiabatic condition is given by
$\epsilon = |d\Omega_k^2/d\tau|/\Omega_k^3 \ll 1 \rightarrow k\tau_1 \gg 1$.
This condition indicates that the first matching occurs in $D2$ or $D4$ in Fig.~1.

(iii) The WKB solution evolves in the intermediate stage,
and enters the attractor stage.
At the attractor stage, it matches with the attractor solution \eqref{muATT}
at $\tau=\tau_2 >\tau_i >\tau_1$.
The coefficients $(A_1',A_2')$ of the attractor solution are then determined
from matching with the WKB solution at $\tau_2$.
Then one can obtain the power spectrum as in Eq.~\eqref{PS}.
At the attractor stage, the adiabatic condition is given by $|k(\tau-\tau_0)| \gg 1$,
which coincides with the subhorizon condition, $\lambda_{\rm phys} \ll H^{-1}$, at this stage.
That means that at the second matching the perturbation is in the subhorizon scale.

(iv) The power spectrum is obtained as a function of $k$
using $(A_1',A_2')$ which are functions of $(c_1,c_2)$,
and evaluated in the superhorizon scale at the end of inflation.
The spectrum is scale invariant for the modes of $k \gtrsim 1/\tau_*$,
and exhibits a peculiar feature for the very long wave-length modes of  $k \lesssim 1/\tau_*$.

\section{Conclusions}
We investigated the tensor perturbation of the inflation model 
in Eddington-inspired Born-Infeld gravity
developed in Ref.~\cite{Cho:2013pea}. 
The background universe is driven by a massive scalar field
as in the usual chaotic inflation model in GR.
There are three stages in the background evolution.
At the early stage of the background evolution,
there is the near-MPS stage which is described by the globally perturbed {\it maximal pressure solution}
investigated in Ref.~\cite{Cho:2013pea}. 
At this stage, the scale factor increases exponentially in cosmological time.
This stage is followed by the intermediate stage. 
For most this stage, the first slow-roll condition is satisfied,
but the analytic form of the background evolution is not available.
At late times, the attractor stage appears in which the first and the second slow-roll conditions are satisfied.
The background evolution is the same with the one in the usual chaotic inflation. 
Therefore, there are two exponentially expanding stages in EiBI inflation.

In EiBI gravity, as one can see from the field equation \eqref{mueq},
the tensor perturbation is described by the conformal time $\tau$
and the scale factor $Y$ of the auxiliary metric $q_{\mu\nu}$ (not of the metric $g_{\mu\nu}$). 
At the attractor stage, $\tau$ is identified with $\eta$ which is the conformal time of the metric. 
The scale factors are related by $Y=Y_0a$, where $Y_0$ is almost constant.
Therefore, the perturbation story  
is very similar to that of the usual chaotic inflation,
with a very small EiBI correction implied in $Y_0$.
The perturbation produced at this stage provides a scale invariant power spectrum.

When the attractor stage does not provide 60 $e$-foldings,
one needs to consider the perturbation produced at the near-MPS stage
in order to explain the low angular modes in the power spectrum.
At the near-MPS stage, $Y(\tau)$ behaves very differently from that at the attractor stage. 
For short wave-length (high $k$) modes, 
the minimum energy condition imposed on the initial perturbation picks 
only the positive energy state. 
It evolves adiabatically at the intermediate stage and is continued to the attractor stage.
The perturbation feature is the same with the one investigated at the attractor stage.

For long wave-length (low $k$) modes, however, the minimum energy condition
requires the mixed energy state of the initial perturbation.
The perturbation can evolve adiabatically at the intermediate stage
for which the WKB solution is applied.
By matching the WKB solution with the near-MPS and the attractor solutions, 
we could evaluate the power spectrum at the attractor stage.
For very low $k$, there is a peculiar rise in the spectrum
while for the rest the spectrum is scale invariant.
This low $k$ behavior may leave a signature in CMB,
which can distinguish the EiBI inflation model from others. 

The recent detection of the $B$-mode polarization by BICEP2 invoked
the importance of the tensor perturbation
in the inflationary scenario \cite{Ade:2014xna}.
According to the result, the tensor-to-scalar ratio is best fit
by the usual $\phi^2$ chaotic inflation model.
Since our EiBI inflation model is very similar to the usual chaotic inflation model
at the attractor stage, we expect that the tensor-to-scalar ratio is also very similar.
We shall investigate the scalar perturbation in the EiBI inflation
in order to confirm this.

\section*{Acknowledgement}
This work was supported by the grants from the National Research Foundation
funded by the Korean government, No. NRF-2012R1A1A2006136 (I.C.) and No. NRF-2013R1A1A2006548 (H.K.).
The authors are grateful to the hospitality of the Asia Pacific Center for Theoretical Physics (APCTP).

\appendix*
\section{Matching time $\tau_1$ and $\tau_2$}\label{app1}
The power spectrum is insensitive to the matching time $\tau_1$ and $\tau_2$.
However, we discuss the constrain on the range of the matching time in this subsection.
At the near-MPS and the attractor stages,
the adiabatic parameter defined in the WKB period is approximated as
\begin{align}
\epsilon_{\rm MPS} &\approx \left[ \frac{3+\kappa\rho}{1+\kappa\rho +6k^2\kappa(1-\kappa p)/a^2}  \right]^{3/2}
\approx \left[ \frac{a^2(3+\kappa\rho)}{6k^2\kappa(1-\kappa p)}  \right]^{3/2}
\quad\mbox{for $k\tau \gg 1$}, \label{EMPS}\\
\epsilon_{\rm ATT} &\approx \left(\frac{\rho a^2/3}{k^2- 2\rho a^2/3} \right)^{3/2}.\label{EATT}
\end{align}
Here, we set $\lambda=1$ for simplicity.

Let us define a new parameter $\varepsilon$ which judges the validity
of the near-MPS and the attractor approximations
by comparing the next order correction in $\Omega_k^2$.
For the near-MPS approximation,
\be
\Omega_k^2 (\tau) = k^2 + \frac{1}{4\tau^2} (1+\psi)
\quad\Rightarrow\quad
\varepsilon_{\rm MPS} = \frac{\psi}{1+4k^2\tau^2} \approx \frac{a^2}{16k^2t_c^2}. \label{EEMPS}
\ee
For the attractor approximations,
\be
\Omega_k^2 (\tau) = k^2 -\frac{2}{3}Va^2 + \frac{\hat\phi^2 a^2}{6(1+\kappa V)}
\quad\Rightarrow\quad
\varepsilon_{\rm ATT} = \frac{\hat\phi^2 a^2}{6(1+\kappa V)(k^2 -2Va^2/3)}
\approx  \frac{\hat\phi^2 a^2}{6(k^2 -2Va^2/3)}, \label{EEATT}
\ee
where we assumed $\kappa V \ll 1$.

We insist that the matching should be performed when the adiabatic parameter is
comparable with the approximation parameter, $\epsilon \approx \varepsilon$.
At the near-MPS stage, the matching time $\tau_1$ is determined
from Eqs.~\eqref{EMPS} and \eqref{EEMPS}  as
\be
\epsilon_{\rm MPS} \approx \varepsilon_{\rm MPS}
\quad\Rightarrow\quad
\tau_1 a^{1/3}(\tau_1) \approx 2k^{-2/3}t_c
\quad\Rightarrow\quad
\tau_1 \approx \left( \frac{16mt_c^3}{3k^2\tau_m}  \right)^{1/(3+m\sqrt{2\kappa/3})},
\ee
if $\kappa\rho \gg 1$.
At the attractor stage, the matching time $\tau_2$ is determined
from Eqs.~\eqref{EATT} and \eqref{EEATT}  as
\be
\epsilon_{\rm ATT} \approx \varepsilon_{\rm ATT}
\quad\Rightarrow\quad
\phi(\tau_2) a^{1/3}(\tau_2) \approx \left( \sqrt{\frac{8}{3}}\frac{k}{m}\right)^{1/3}
\quad\Rightarrow\quad
\tau_2 \approx \tau_0 +\sqrt{\frac{3}{8}}\frac{\phi_i^3ma_i(\tau_i-\tau_0)}{k},
\ee
where the last approximation valid only when $\phi(\tau_2) \approx \phi_i$.
(Otherwise, $\tau_2$ is obtained only implicitly.)

\end{document}